\documentclass[twocolumn,showpacs,prb,10pt]{revtex4}
\usepackage{graphicx}

\begin{document}

\title {Thermal transport in a spin-1/2 Heisenberg chain \\ 
coupled to a (non) magnetic impurity}
\author{A. Metavitsiadis$^1$ and X. Zotos$^1$}
\affiliation{$^1$ Department of Physics, University of Crete and
Foundation for Research and Technology-Hellas, P.O. Box 2208, 71003
Heraklion, Greece}
\author{O. S. Bari\v si\'c$^{2,3}$ and P. Prelov\v sek$^{3,4}$}
\affiliation{$^2$ Institute of Physics, HR-10000 Zagreb, Croatia}
\affiliation{$^3$J.\ Stefan Institute,
SI-1000 Ljubljana, Slovenia}
\affiliation{$^4$ Faculty of Mathematics
and Physics, University of Ljubljana, SI-1000 Ljubljana, Slovenia}

\date{\today}

\begin{abstract}
We explore the effect of a (non) magnetic impurity on the thermal
transport of the spin-$1/2$ Heisenberg chain model. This unique system
allows to probe Kondo-type phenomena in a prototype strongly
correlated system.  Using numerical diagonalization techniques we
study the scaling of the frequency dependent thermal
conductivity with system size and host-impurity coupling strength as well 
as  the dependence on temperature. 
We focus in particular on the analysis of
``cutting-healing" of weak links or a magnetic impurity
by the host chain via Kondo-like screening as the temperature is
lowered.
\end{abstract}

\maketitle

\section{Introduction}

A (non) magnetic impurity coupled to a spin-$1/2$ Heisenberg chain is
a prototype system that exemplifies ``Kondo"-type effects in a
correlated system.  Starting with the proposal of
Kane-Fisher,\cite{kf} a weak link in a repulsive (attractive)
Luttinger liquid was shown to lead to an insulating (transmitting)
ground state.  The cutting or healing of spin chains by a variety of
(non) magnetic defects has also been
established\cite{affleck,eggert,sorensen,laflorencie} as well as the
effect of a magnetic impurity on the ground state of the anisotropic
easy-plane Heisenberg chain.\cite{furusaki} Generically, a weak link
or coupling to a magnetic impurity in a Heisenberg antiferromagnetic
chain leads to a ground state corresponding to two open chains. In the
exceptional case of two adjacent links or a ferromagnetic (attractive in the 
fermionic language) easy axis anisotropy a healing of the defect is
conjectured.\cite{furusaki}  This screening effect is characterized by
a Kondo-like temperature and screening
length.\cite{eggert,sorensen}  These phenomena have so far mostly been
studied either as they are reflected on ground state properties,
e.g. finite size gaps, entanglement or, somewhat indirectly, as a
temperature dependent induced staggered susceptibility.\cite{eggert}

In this work we use an exceptional physical probe for the study of
these effects, namely, the thermal transport in the spin-$1/2$
Heisenberg chain that is truly singular.  Although the Heisenberg
model describes a strongly correlated system, the thermal conductivity
is purely ballistic as the energy current commutes with the
Hamiltonian,\cite{znp} a result that is related to the integrability
of this model.\cite{higherspin}  Thus the only scattering present is
due to the defect and thus its frequency/temperature/coupling strength
dependence can be isolated and clearly analyzed. In this context it
was already found that a single potential impurity renders the thermal
transport incoherent\cite{static} with the frequency dependence of the
thermal conductivity well described by a Lorentzian, at least for
  a weak impurity.  This is in sharp contrast to the case of a
non-interacting system where in spite of the impurity the
transport remains coherent described within the Landauer formalism by
a finite transmission coefficient through the impurity.  Thus a single static
impurity materializes the many-body character of scattering states.

Besides its theoretical interest, the effect of (non)magnetic impurities on 
the thermal transport of quasi-one dimensional materials as SrCuO$_2$,  
Sr$_2$CuO$_3$ and the ladder compound La$_5$Ca$_9$Cu$_{24}$O$_{41}$ has 
recently become possible to explore experimentally.\cite{hess}

In this work we use numerical diagonalization techniques - 
(full) exact diagonalization (ED), the Finite-Temperature Lanczos
method (FTLM)\cite{ftlm} and the Microcanonical Lanczos method
(MCLM)\cite{mclm} - to study the thermal transport in the
Heisenberg chain model either coupled to a magnetic impurity
or perturbed by single and double weak links.  These state of
the art techniques are crucial in the attempt to look for the subtle
low temperature many-body effects associated with Kondo screening.

\section{Model}
We consider the one dimensional anisotropic spin-$1/2$ Heisenberg model 
in the presence of a magnetic impurity out of the chain or weak links, 
\begin{eqnarray}
H&=&\sum_{l=0}^{L-1} J_{l,l+1} h_{l,l+1}+J'
(s^x_0S^x+s^y_0S^y+\Delta' s^z_0S^z),
\nonumber\\
h_{l,l+1}&=&s^x_ls^x_{l+1}+s^y_ls^y_{l+1}+\Delta
s^z_ls^z_{l+1},
\label{ham}
\end{eqnarray}
where $s^{\alpha}, \alpha=x,y,z$ are spin-$1/2$ operators, $J_{l,l+1}>0 $ 
the in-chain magnetic exchange coupling that we take antiferromagnetic, 
$J'$ the chain-impurity coupling, 
$\Delta, \Delta'$ anisotropy parameters and {\bf S} a spin-S 
magnetic-impurity operator ($\hbar=1$).
In this work we mostly consider a spin-$1/2$ impurity. We 
assume periodic boundary conditions, ${\bf s}_{L}={\bf s}_{0}$, 
and uniform couplings $J_{l,l+1}=J$, except in 
the study of weak links (see below).
We vary the anisotropy parameters $\Delta, \Delta'$, 
with $\Delta=\Delta'$, in order to 
look for the (healing) cutting of the chain effects mentioned above.

In our study, based on standard linear response theory,
the frequency $\omega$ dependence of the real part 
of the thermal conductivity (regular component) is given by 
\begin{equation}
\kappa(\omega)=-\frac{\beta}{\omega}\chi''(\omega),~~ 
\chi(\omega)=
\frac{i}{L}\int_0^{+\infty} dt e^{i\omega t}
\langle[j^{\epsilon}(t),j^{\epsilon}]\rangle,
\end{equation}
where $\beta=1/T$, $T$ is the temperature and  $k_B=1$.
We determine the energy current from the hydrodynamic ($q\rightarrow 0$) 
limit of the energy continuity equation 
$\partial H_q/\partial t\sim q j^{\epsilon}$ 
with $H_q=\sum_l e^{iql} h_{l,l+1}$ as, 

\begin{eqnarray}
j^{\epsilon}&=&
%J^2 
\sum_{l=0}^{L-1} J_{l-1,l}J_{l,l+1}\: {\bf s}_l\cdot( {\bf s}_{l+1}\times{\bf
  s}_{l-1})\nonumber\\ &+&\frac{JJ'}{2}{\bf s}_0\cdot( {\bf
  S}\times{\bf s}_{L-1}+ {\bf s}_1\times{\bf S}),
\label{je}
\end{eqnarray}

\noindent
showing for simplicity the case $\Delta=1$ ($\Delta\ne 1$ is obtained by
$s_l^z \rightarrow \Delta s_l^z$ in the cross-product terms).  When
$J'=0$ and all $J_{l,l+1}=J$ the energy current commutes with the Hamiltonian, 
the transport
is purely ballistic and the thermal conductivity consists of only a
$\delta(\omega)$-peak proportional to the thermal Drude weight.
\section{High temperature limit}

Starting from the high temperature ($\beta\rightarrow 0$) limit 
we can obtain a first impression on the behavior of the frequency 
dependence of $\kappa(\omega)$ from the 0th and 2nd moments, 
$\mu_n=\int d\omega \omega^n \kappa(\omega)$ 
%which are proportional to, 
%
%\begin{equation}
%\mu_0\sim J^2,~~~ \mu_2\sim \frac{J^4{\cal B}^2}{L}
%\label{mu}
%\end{equation}
%
which are equal to (for the isotropic point, $\Delta=1$), 
\begin{eqnarray}
\mu_0&=&\text{const.}\times\frac{6}{T^2}
\big(J^2+\frac{2}{L}\mathcal{B} ^2\big),\quad
\big(\text{const.}=\pi\frac{J^2}{64}\big)\label{mu}\\
\mu_2&=& \text{const.}\times \frac{\mathcal{B}^2}{LT^2}
\left(39J^2-12JJ'+3J'^2+36\mathcal{B}^2\right)\:,\nonumber
\end{eqnarray}
\noindent
where ${\cal B}^2=(J'^{\:2}/3)S(S+1)$ is the characteristic impurity
spin dependence. One could expect the 2nd moment to reflect the
width of $\kappa(\omega)$ and thus to be related to the inverse
scattering time $1/\tau$.  We note that for this impurity problem 
an assumption of a Gaussian form
$\kappa(\omega)=\kappa_{dc}\exp^{-(\omega\tau)^2}$ would imply from
the $L$ dependence of $\mu_{0,2}$ that $\kappa_{dc}=\kappa(0)$ would
scale as $\sqrt{L}$ and $1/\tau \sim 1/\sqrt{L}$. This is,
however, incorrect as is also evident from the disagreement with higher
moments, $n>2$, which behave all as $\mu_n \propto 1/L$.  For 
weak-coupling cases, such as a single impurity weakly coupled to the host 
chain, we should therefore
rather expect a Lorentzian-like frequency dependence with a
static $\kappa(0) \propto L$ and a characteristic frequency width
$1/\tau \propto 1/L$.

\begin{figure}[ht]
\includegraphics[angle=0, width=.9\linewidth]{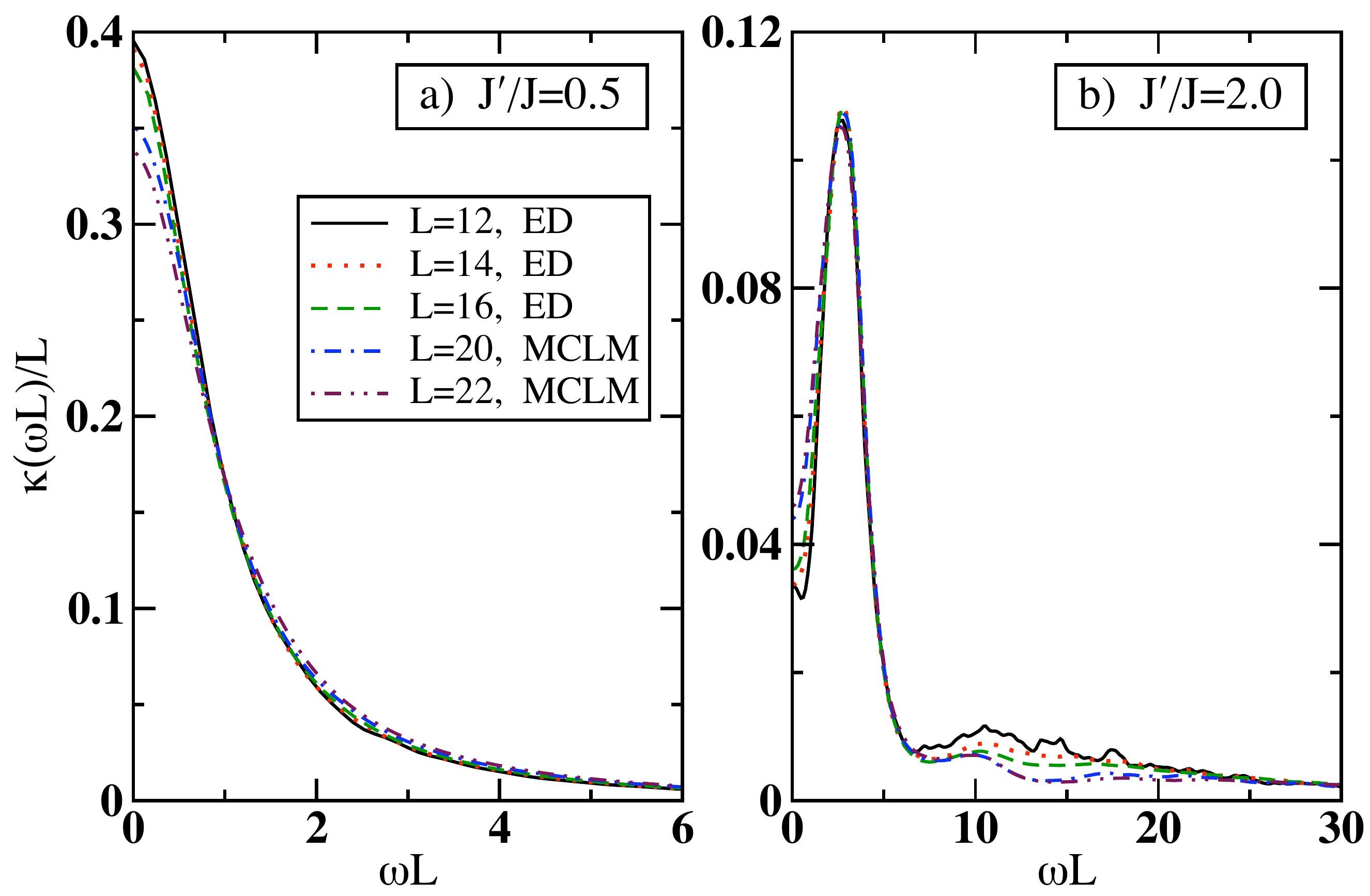}
\caption{Frequency-dependent thermal conductivity in the high-$T$
limit scaled as $\kappa(\omega L)/L$ for ($\Delta=1$): 
(a) weak coupling $J'=0.5J$, (b) strong coupling
$J'=2J$ (curves are normalized to unity).}
\label{l_scale}
\end{figure}

In Fig. \ref{l_scale} we show the frequency dependence of the
thermal conductivity, normalized and appropriately scaled with system size. 
Note that in the high-$T$ ($\beta
\rightarrow 0$) limit the relevant (but still nontrivial) quantity
is $T^2 \kappa(\omega)$ which is implicitely extracted by the normalization. 
We thus present results of the normalized $\kappa(\omega
L)/L$ for a weak, $J'=0.5J$ and strong, $J'=2J$ coupling case
respectively.  The data up to $L=16$ were obtained by full ED
while for $L=18-22$ the MCLM was used.\cite{mclm}  The $\delta-$peaks at
the excitation frequencies are binned in windows $\delta\omega=0.01$,
which also gives the frequency resolution of the spectra.  For
$J'=0.5J$ we find a simple Lorentzian form while in the strong
coupling case the behavior is nonmonotonic with a maximum at a finite
frequency $O(1/L)$.  In both cases the proposed $L$ scaling is indeed
realized.
\begin{figure}[ht]
\includegraphics[angle=0, width=.9\linewidth]{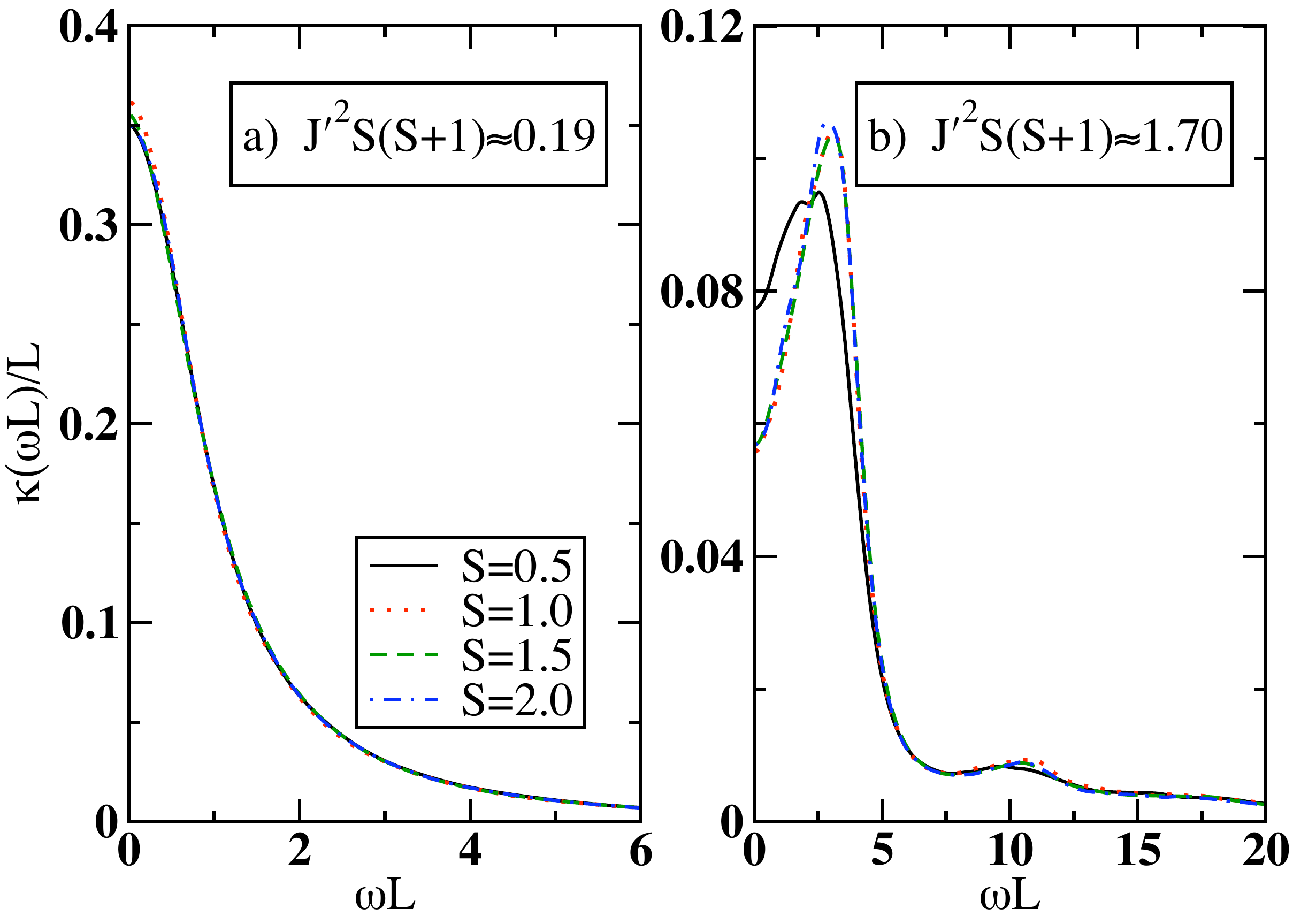}
\caption{Frequency dependence of the normalized thermal conductivity
$\kappa(\omega L)/L$ in the high-$T$ limit for a variety of
impurity spin values $S=1/2,1,3/2,2$ and for: (a) 
$J'/J=0.5,0.3,0.22,0.18$ corresponding to the weak coupling 
${\cal B}^2=(J'^{\:2}/3)S(S+1)\simeq0.06$, (b)
$J'/J=1.5,0.92,0.67,0.53$ corresponding to the stronger coupling ${\cal
B}^2\simeq 0.57$.}
\label{s_scale}
\end{figure}

As for the scaling with impurity spin $S$ suggested by the
proportionality of the 2nd moment to ${\cal B}^2=(J'^{\:2}/3)S(S+1)$
we show in Fig. \ref{s_scale} MCLM results for $\kappa(\omega
L)/L$ for a series of S-values and couplings $J'$ so that the effective
perturbation strength ${\cal B}^2$ retains its value.  We
find indeed that at both weak as well as strong coupling the scaling
is well obeyed, giving a wider applicability to our results. 
They can be applied to a range of impurity spin values becoming directly 
relevant in the interpretation of experiments.

\begin{figure}[ht]
\includegraphics[angle=0, width=.9\linewidth]{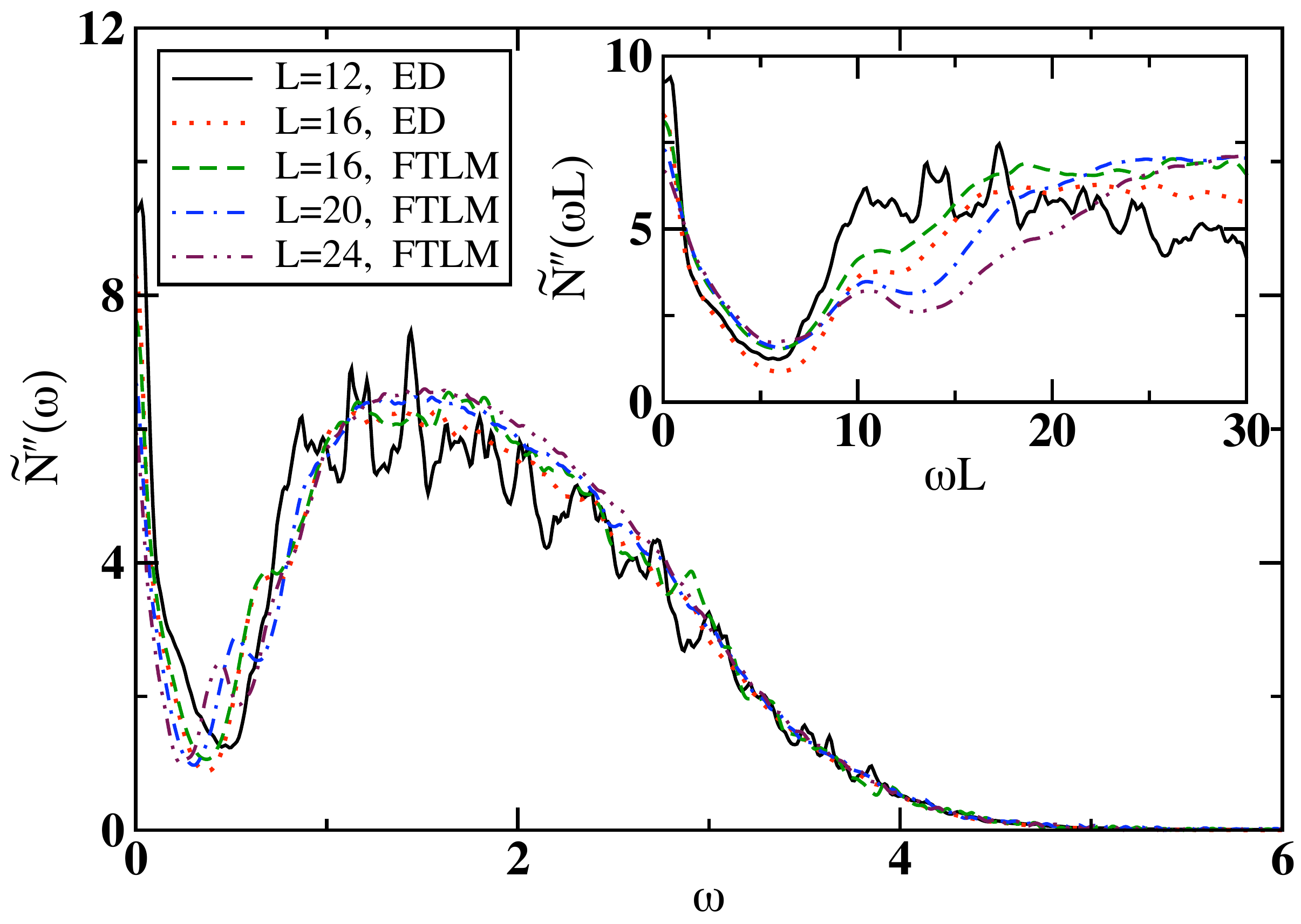}
\caption{Memory function $\tilde N''(\omega)$ for a strong coupling
$J'=2J$ and for various lattice sizes $L=12-24$, using both ED and
FTLM. Inset: the scaled function $\tilde N''(\omega L)$ is shown at
low frequencies.}
\label{w_scale}
\end{figure}
Now let us address the generic $L\rightarrow \infty$ behavior. We can 
discuss it by considering the memory function $N(\omega)$ representation
defined via the general complex function $\bar{\kappa}(\omega)$,
\begin{eqnarray}
\bar{\kappa}(\omega)&=&i\beta\frac{\chi_0} {\omega+N(\omega)},~~~
\chi_0= \chi(\omega\rightarrow 0).
\end{eqnarray}
where the real $\kappa(\omega)=\bar{\kappa}'(\omega)$ and
$N''(\omega) \sim 1/\tau$ plays the role of the (frequency
dependent) thermal-current relaxation rate. The lowest moments $\mu_n$
can be evaluated (in principle) exactly in the high-$T$
limit\cite{ins} on a finite size lattice of $L$ sites. Involving
only local quantities, at least for $0<n<L/2$, they should behave as
$\mu_n = \tilde \mu_n/L$ whereby $\tilde \mu_n$ is size independent
for $n<L/2$.  It is plausible that also higher moments, $n>L/2$, behave
as $\mu_n \propto 1/L$. 
%If also the $\tilde \mu_n$ were size independent 
 If $\tilde \mu_n$ for $n>L/2$ would be also size independent,
then  this would imply the scaling
$N(\omega)=\frac{1}{L}\tilde N(\omega)$, with a universal (size
independent) $\tilde N(\omega)$. Consequently
\begin{equation}
\bar{\kappa}(\omega)=\frac{i\beta\chi_0 L}{(\omega L) + \tilde N (\omega)},  
\end{equation}
with the real part $\kappa(\omega)$ for $L\rightarrow \infty$ and
$\omega \to 0$ obeying 
the Lorentzian scaling relation,
\begin{equation}
\frac{\kappa(\omega L)}{L}=\frac{\beta\chi_0 \tilde N''(\omega\rightarrow 0)}
{(\omega L)^2 + \tilde N''(\omega\rightarrow 0)^2},
\end{equation}
provided that $N''(\omega\rightarrow 0)$ is finite. This is, however,
clearly not what we observe in Fig. \ref{l_scale}, where from the
non-Lorentzian shape we must conclude that the memory function also scales
as $\tilde N(\omega L)$ and thus,
\begin{equation}
\frac{\kappa(\omega L)}{L}=\frac{\beta\chi_0 \tilde N''(\omega L)}
{(\omega L + \tilde N' (\omega L) )^2 + \tilde N''(\omega L)^2}.
\end{equation}
This is not in contradiction with the moments argument, since the 
higher moments, $n>L/2$, determine the low frequency behavior. 
So we can argue that
at high frequencies $\tilde N(\omega)$ scales as $\omega$ while at low
frequencies as $\omega L$. This scenario is indeed verified in
Fig. \ref{w_scale} at the low/high frequency regimes, where
$N(\omega)$ is extracted from the $\kappa(\omega)$ data. 
The FTLM method is used for lattice sizes $L\ge16$ with  
$M_L=500$  Lanczos steps and smoothed with an additional frequency 
broadening $\delta\omega=0.03$.
On the other hand, we can also explain the observed general $\kappa(\omega
L)/L$ scaling with the similarity to a noninteracting system - with
an impurity. In the latter case, the characteristic scaling $L \omega$
is signature of ``free'' oscillations in the system.
\begin{figure}[ht]
\includegraphics[angle=0, width=.9\linewidth]{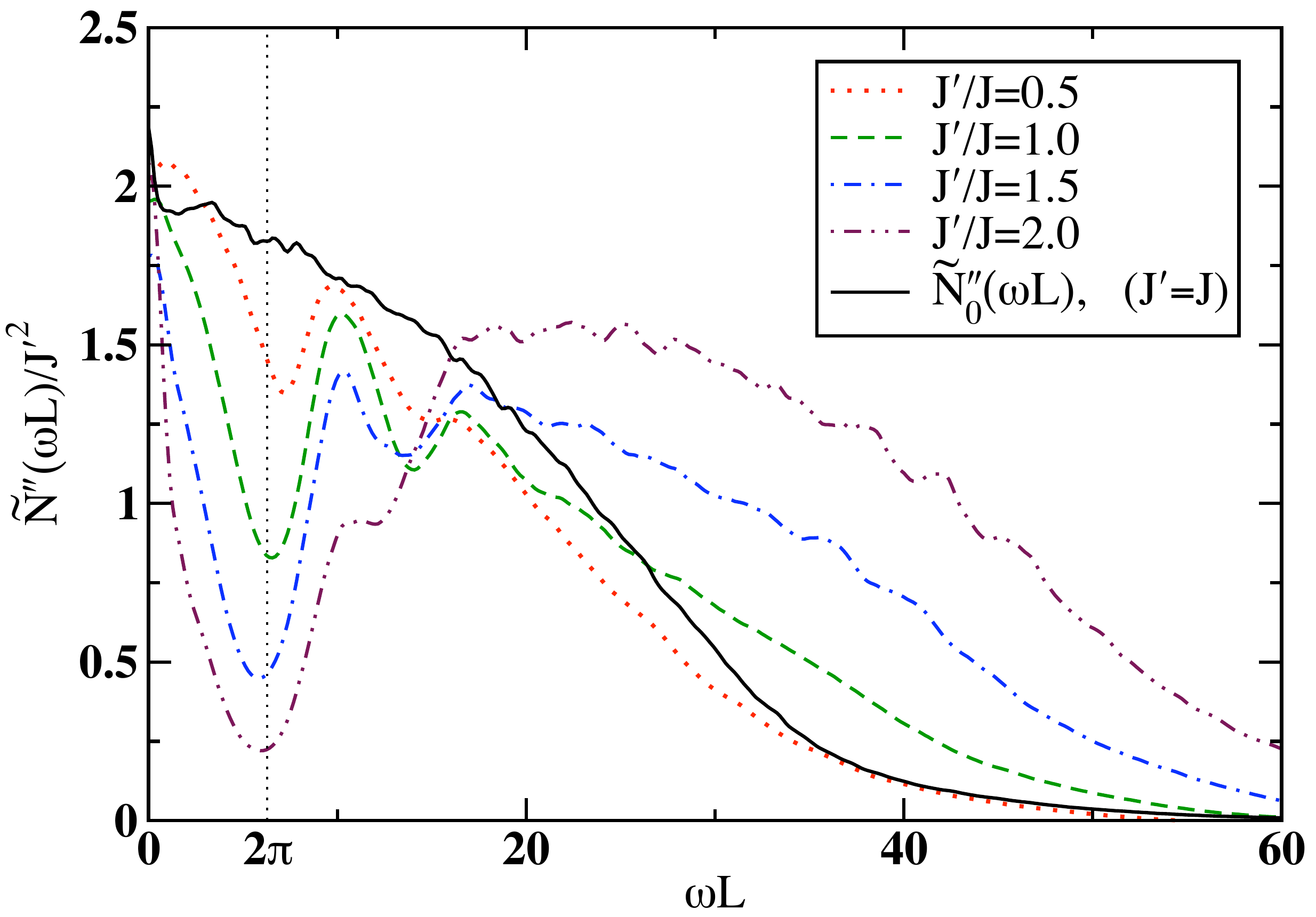}
\caption{Impurity coupling $J'$ dependence of scaled $\tilde N''(L
\omega)/J'^2$ and the comparison with the perturbative
result. Results are obtained for $\Delta=1$ and $L=16$ via ED.}
\label{j_scale}
\end{figure}

To study the crossover from weak to strong coupling regime we show in
Fig. \ref{j_scale} the evolution of the relaxation-rate function
$\tilde N''(\omega L)$ with impurity coupling $J'$ along with a
perturbative evaluation $\tilde N_0''(\omega L)$ using the eigestates
of the Hamiltonian without the impurity.\cite{gw}  It is interesting
that the memory function shows an increasingly pronounced structure
with minima at approximately the same frequencies, multiples of
$2\pi/L$ {\it independently of $J'$} and which are not present in the
perturbative calculation. In particular the characteristic frequency 
of the minima decreases as the anisotropy parameter $\Delta$ decreases 
and thus it apparently related to the velocity of elementary excitations 
(spinons) in the system. We can conjecture that this peak structure
is due to a resonant mode, created by 
%backscattering to a chain -impurity bound state. 
multiple forward/backward scattering on the impurity, 
characteristic for the noninteracting system.
It is remarkable that this happens even in this
high temperature limit. This effect has already been seen in
integrable systems where a perturbation seems to affect the totality 
of the energy spectrum.\cite{ins} 
Now the picture is clear, $\tilde N''(\omega)$
increases as $J'^2$, scales as $\omega L$ at low frequencies and at
the same time develops a structure that dominates the behavior of
$\kappa(\omega L)$ turning the Lorentzian weak-coupling shape to a
nontrivial one at strong coupling.

\section{Weak links - finite $T$}

Next we examine the behavior of the thermal conductivity
$\kappa(\omega)$  as we lower the temperature, starting with the
influence of static weak exchange links.

Kane-Fisher\cite{kf}  for a Luttinger liquid and
Eggert and Affleck\cite{affleck} (EA) for the isotropic
spin-$1/2$ Heisenberg chain, proposed that a weak link
leads to an open chain (cutting) in the low energy
limit. In contrast, a defect of two adjacent weak links is ``healed"
leading to a uniform chain at $T=0$.

\begin{figure}[ht]
\includegraphics[angle=0, width=.9\linewidth]{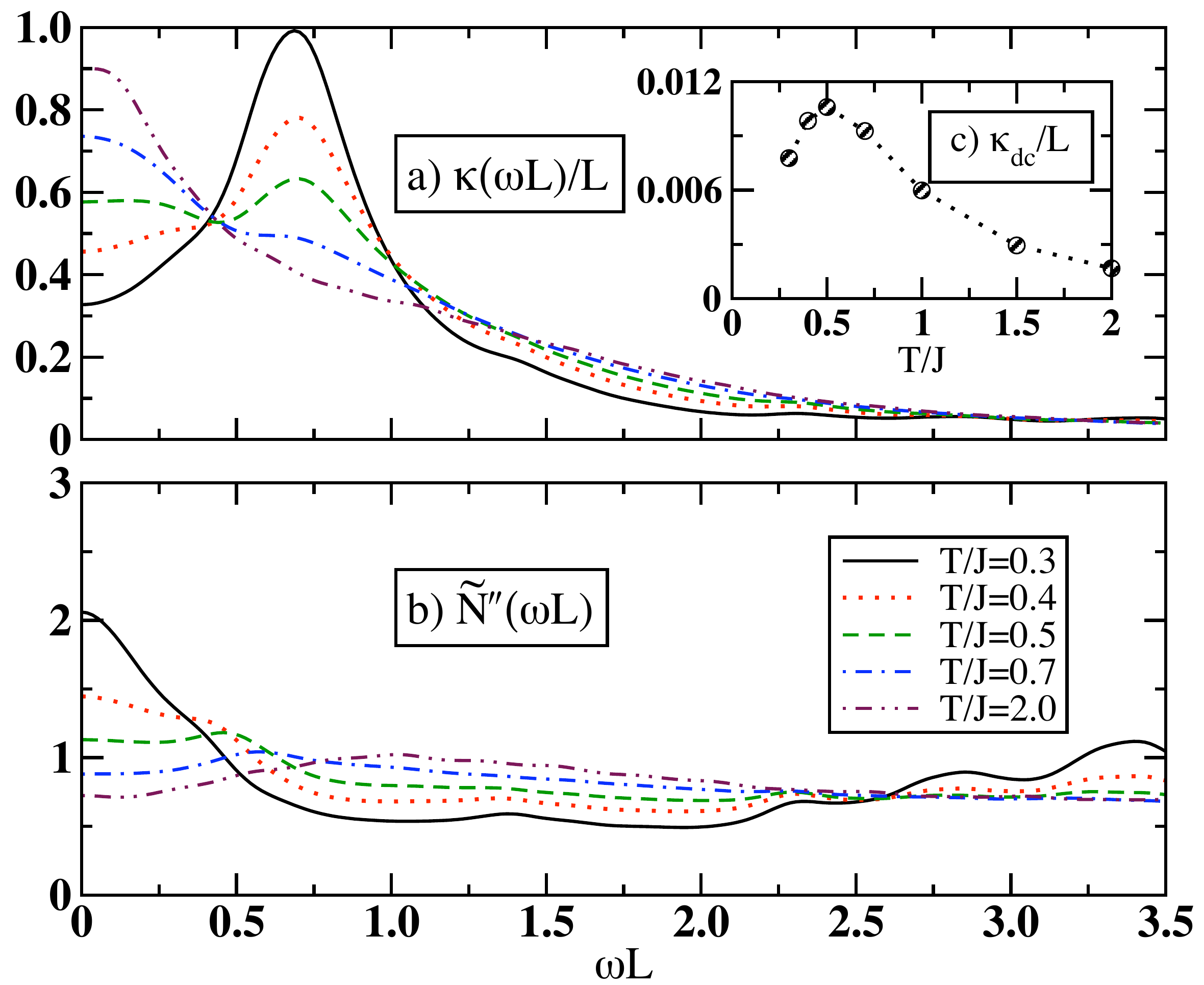}
\caption{Frequency dependence of: (a) the normalized thermal
conductivity $\kappa(\omega L)/L$, (b) the extracted memory function
$\tilde N''(\omega L)$, for a chain of $L=22$ sites with one weak link
$\tilde J=0.7J$ and various $T/J=0.3 - 2.0$ .  (c) Temperature
dependence of $\kappa_{dc}(T)/L$.}
\label{wl1}
\end{figure}

To analyze this effect we consider a chain with only one weak link, that
is one altered bond with coupling e.g. $J_{0,1}=\tilde J$ in an
otherwise uniform chain ($J'=0$, there is no spin impurity).  
The characteristic Kane-Fisher temperature is given in the weak coupling
limit by $T_{KF}\sim (J-\tilde J)^2/J$.
In Fig. \ref{wl1}a we show the corresponding $\kappa(\omega L)/L$ for
$\tilde J=0.7J$ and a series of temperatures. The data are obtained
using the FTLM method for a chain of $L=22$ spins, by $M_L=2000$
Lanczos steps and smoothed by an additional frequency broadening
$\delta \omega = 0.007$. From Fig.~\ref{wl1}a 
we notice that $\kappa(\omega L)/L$
develops a strongly nonmonotonic frequency dependence by lowering the
temperature, with a maximum at a finite frequency that suggests a flow
to the strong coupling limit similar to the one discussed before by
increasing $J'$.  In Fig.~\ref{wl1}b, the extracted $\tilde N''(\omega L)$
for various $T$ is presented, with the development of a
characteristic structure that explains the nonmonotonic behavior of
$\kappa(\omega)$.  The increasing value of $\tilde N''(0) \sim 1/\tau$
with decreasing temperature indeed corresponds to the effect of
``cutting" of the chain.

\begin{figure}[ht]
\includegraphics[angle=0, width=.9\linewidth]{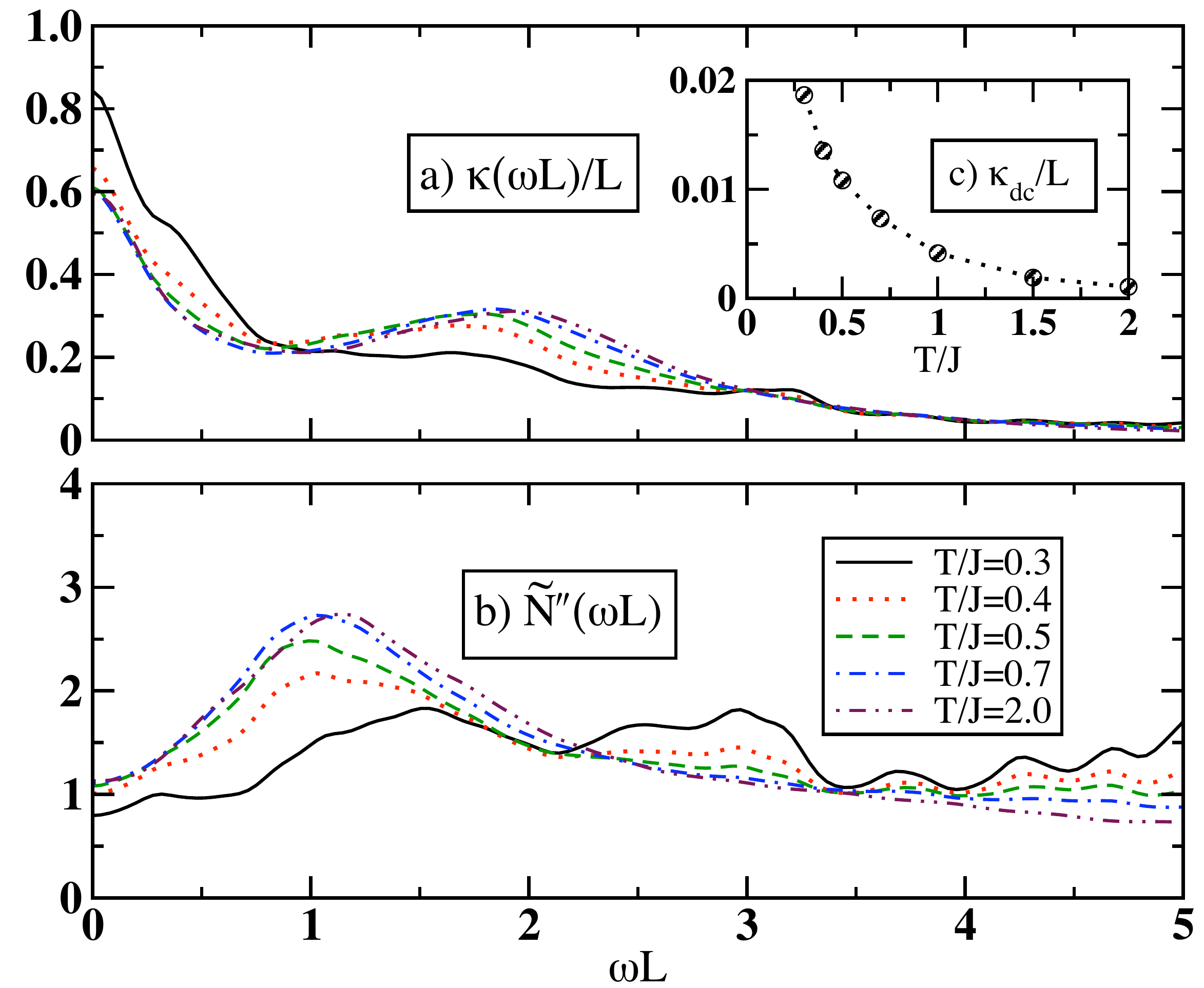}
\caption{Frequency dependence of: (a) the normalized thermal conductivity
$\kappa(\omega L)/L$, 
(b) the extracted memory function  $\tilde N''(\omega L)$  
for a chain of $L=22$ sites with two adjacent weak links $\tilde J=0.7 J$ 
and various $T/J=0.3 - 2.0$. (c) Temperature dependence of $\kappa_{dc}(T)/L$.}
\label{wl2}
\end{figure}

Nonmonotonic is also the frequency dependence of 
$\kappa(\omega L)/L$  for the case of two  adjacent equal 
weaker links, $J_{L-1,0}=J_{0,1}=\tilde J= 0.7 J$, as shown in 
Fig. \ref{wl2}a. However, in this case we observe in Fig. \ref{wl2}b
the opposite behavior of $\tilde N''(\omega)$. Namely ``healing" of the 
double defect deduced by the decreasing 
$\tilde N''(0)$ as the temperature is lowered in agreement with theoretical 
prediction.\cite{affleck} We should note that both cutting/healing are 
low frequency effects at frequencies $\omega L\:\: O(1)$.

\begin{figure}[ht]
\includegraphics[angle=0, width=.9\linewidth]{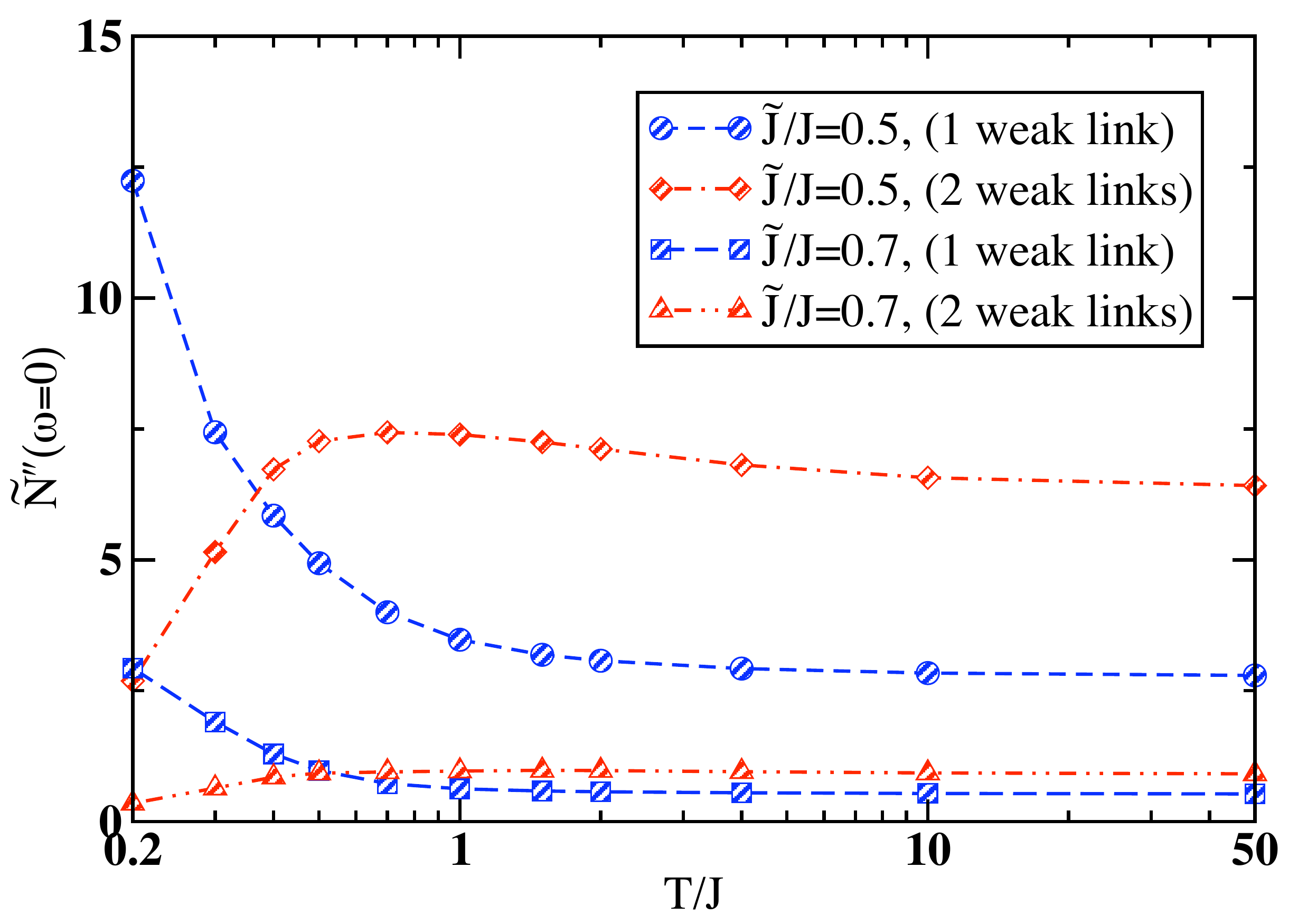}
\caption{Temperature dependence of $\tilde N''(0)$ for $\tilde{J}=0.5,0.7J$ 
showing cutting/healing behavior for one and two weak links.}
\label{wl12n0}
\end{figure}
To summarize the observed behavior we show in Fig. \ref{wl12n0},
the $T$-dependence of the relaxation rate $\tilde N''(0)$  
for two different couplings $\tilde{J}/J=0.5,0.7$, 
for one and two weak links, respectively. The presented results confirm the 
existence of the cutting behavior at low $T$ for a single link, as well 
as the healing by lowering $T$ for two adjacent and equal links. 
As expected, both effects appear only at low $T/J < 1$ while the dependence 
of the characteristic $T_{KF}$ on $\tilde J/J$ is less pronounced.

\section{Spin coupled to the chain - finite $T$}

Finally we can study the effect of lowering the temperature on the 
scattering by a magnetic impurity.
According to EA it leads to cutting the chain at  $T=0$ 
irrespective of the sign of $J'$.
This proposal was extended by Furusaki and Hikihara\cite{furusaki} 
to the anisotropic spin chain $-1 < \Delta \le 1$ where they 
furthermore proposed that for $-1 < \Delta < 0$ (attractive case in 
the fermionic language) there is ``healing" of the impurity, in analogy 
to the case of two adjacent weak links.

In the Kondo problem the characteristic temperature in the 
weak coupling limit is given by $T_K\sim v \exp(-c/J')$ with 
$c$ being a constant, $v$ the velocity 
of spin excitations and $J'$ the Kondo coupling. In the case of a spin-$1/2$ 
chain it was shown\cite{frahm} that the exponential dependence is 
replaced by $T_K\sim \exp (-\pi \sqrt {1/J' -(S'+1/2)^2})$ 
and a next-nearest neighbor coupling 
$J_2\simeq 0.2412$ is needed to recover the traditional Kondo case.
We should note that in the model studied the impurity spin is attached 
only at the end of the chain  - in contrast to our model - but plausibly 
the behavior is qualitatively similar. To get a qualitative idea of 
orders of magnitude for our problem \cite{laflorencie} for 
$J'=0.3J$, $T_K\sim 0.014$, $\xi_K\sim 40$,  
for $J'=0.6J$, $T_K\sim 0.388$, $\xi_K\sim 4$ and $J'=J$, $\xi_K=0.65$. 
As in our study we are limited to $T \ge 0.4$ in order to see a ``Kondo" 
crossover we must consider a coupling 
$J' \ge 0.5J$ and thus we are in the relatively strong coupling regime,
with typical screening length of the order $\xi_K \sim 1$. 
\begin{figure}[ht]
\includegraphics[angle=0, width=.9\linewidth]{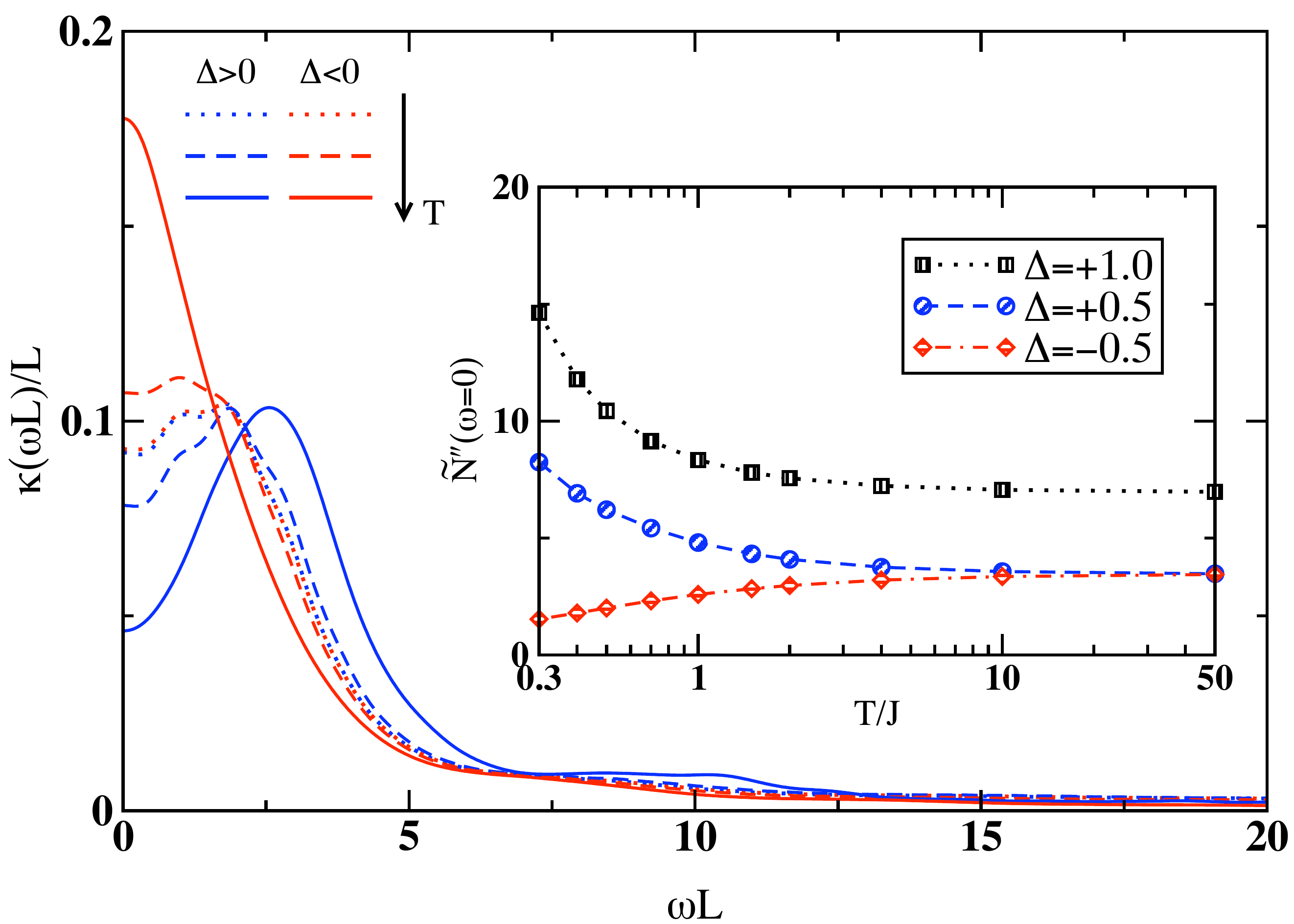}
\caption{Frequency dependent normalized thermal conductivity 
$\kappa(\omega L)/L$ for  strong coupling $J'=2J$, $\Delta=\pm0.5$ and 
three  $T/J=50,2,0.4$. Inset: $T$-dependence of $N''(0)$ for $\Delta=\pm0.5$ 
and $\Delta=+1$.}
\label{cutheal}
\end{figure}

In Fig. \ref{cutheal} we show $\kappa(\omega L)/L$ for a chain of $L=22$ sites 
at strong coupling $J'=2J$ and two representative cases $\Delta=\pm 0.5$ 
as we lower the temperature.
Indeed we find at low frequencies the gradual development of the corresponding 
``cutting/healing" behavior which we exemplify in the inset by $\tilde N''(0)$ 
as a function of temperature both for $\Delta=\pm 0.5$ and the 
most typical isotropic case $\Delta=+1.0$.
It is remarkable that the tendency to increase-decrease the scattering time 
is already evident from high $T$, presumably due to the local 
character of the effect because of the strong $J'$ coupling. 
We note in passing that the $\omega L$ scaling is found 
not just at high $T$ but rather at all $T$ (not shown).

\begin{figure}[ht]
\includegraphics[angle=0, width=.9\linewidth]{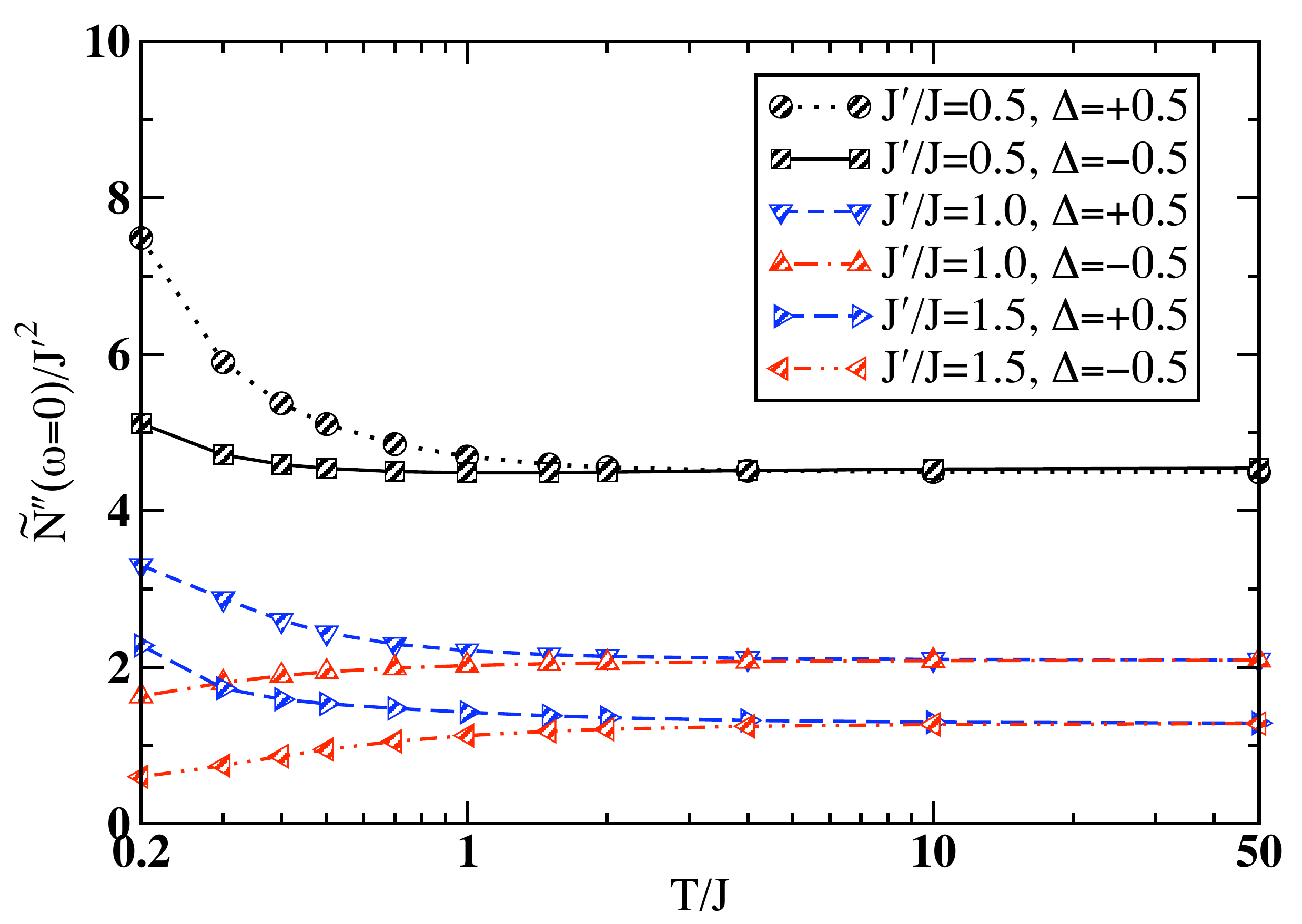}
\caption{$\tilde N''(0)$ vs. $T$ for the 
repulsive (attractive) case $\Delta=+0.5 (-0.5)$  for different
$J'/J=0.5,1.0,1.5$.}
\label{memjp}
\end{figure}

Next in Fig. \ref{memjp} we show $\tilde N''(0)$ as a function of $T$ 
for a series of increasing $J'$ couplings. The ``cutting" effect for 
the repulsive case $\Delta=+0.5$ is present for all values of $J'$ 
with no easily distiguishable ``Kondo" temperature. We are always 
dealing with screening lengths well less than the system size where presumably 
no subtle many-body effects come into play. On the other hand, 
in the attractive case 
$\Delta=-0.5$, we do not observe ``healing" for the weakest coupling 
$J'=+0.5$ where the screening length is expected to be several lattice sites.

\begin{figure}[ht]
\includegraphics[angle=0, width=.9\linewidth]{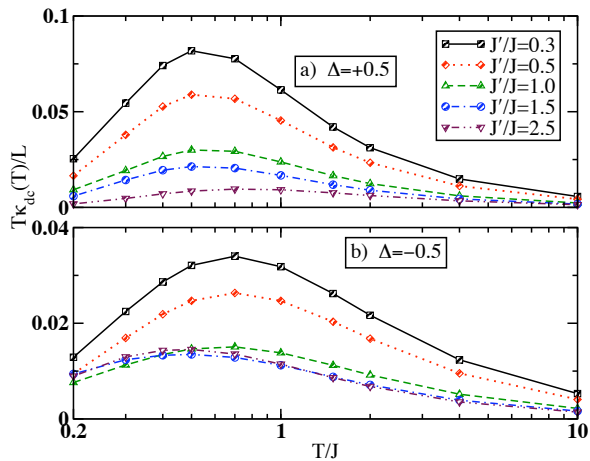}
\caption{Temperature dependence of $\kappa_{dc}/L$ for a variety of 
impurity couplings $J'$ and for: (a) repulsive $\Delta=+0.5$, 
(b) attractive $\Delta=-0.5$.}
\label{kdc}
\end{figure}

\begin{figure}
\includegraphics[angle=0, width=.9\linewidth]{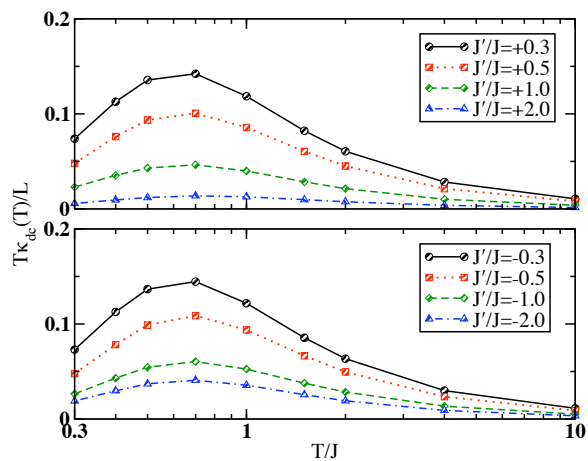}
\caption{Temperature dependence of $\kappa_{dc}/L$ for a variety of 
impurity couplings $J'$, $\Delta=1$ and for:  anti-ferromagnetic
couplings (top), ferromagnetic couplings (bottom).}
\label{kdciso}
\end{figure}

Finally, in Fig. \ref{kdc} we summarize the $T$-dependence of $\kappa_{dc}/L$ 
for a variety of coupling strenghts $J'/J$ and 
$\Delta=\pm 0.5$. The experimentally most interesting 
case $\Delta=+1$ corresponding to isotropic antiferromagnetic 
as well as ferromagnetic 
impurity coupling is shown in Fig. \ref{kdciso}. 
For $\Delta>0$ we observe in Fig. \ref{kdc}a and
Fig. \ref{kdciso} a continuous decrease of the $\kappa_{dc}$ 
with increasing $J'$. This can be explained
with the formation of a local singlet, at least for $T<J'$ 
which blocks the current through the impurity region. 
On the other hand, the $\Delta<0$ case in Fig. \ref{kdc}b
reveals a saturation of $\kappa_{dc}$ with $J'$, 
at least for intermediate large $J'$. 
However, for severe perturbations ($J'\gg J$)  the impurity cannot be 
healed by the chain leading inevitably to a further decrease of the $\kappa_{dc}$.


\section{Conclusions}

In conclusion, by analysing the unique behavior of the thermal conductivity 
of the spin-1/2 Heisenberg model 
several effects of the local static and dynamical impurities have been established:
\\(\textit{a})
A single local impurity, either static as the local field \cite{static} and
 weak link, or dynamical as the spin coupled to the chain turn the
dissipationless thermal conductivity into an incoherent one. 
Numerical results for the dynamical  conductivity,  best studied at high-$T$, 
reveal that a single impurity in a system of $L$ sites shows a universal 
scaling form $\kappa(\omega L)/L$ at least in the low-$\omega$ regime.
For weak perturbation, as weakly coupled spins outside 
the chain, the scaling form is of the simple Lorentzian type. 
On the contrary large local perturbation can  lead to a nontrivial 
form with the  maximum response at $\omega >0$.
%
%\item[b)]
\\(\textit{b})
Furthermore,  universal oscillations in the dynamical
relaxation rate $N''(\omega)$ become visible, from 
the weak coupling regime already,  with the period 
$\omega \propto 1/L$  being a  remnant of the impurity 
multiple-scattering phenomena in a noninteracting system.
%
%\item[c)] 
\\(\textit{c})
Our results confirm the existence of the Kondo-type effects of
impurities on  lowering the temperature. In the case of weak links 
and for the isotropic Heisenberg model cutting and healing  
effects are  observed   at lower $T$ for a single weak link 
and a pair of identical weaker links,  respectively, in accordance with theoretical predictions.\cite{kf, eggert}  In the case of a spin coupled to the chain the  
cutting/healing effects at  low $T$ depend  on the sign of the anisotropy 
$\Delta$. For ferromagnetic anisotropy ($\Delta<0$), the chain
screens the impurity and the system enters the weak coupling regime 
as the temperature is decreased. The opposite behavior is obtained 
for antiferromagnetic anisotropy ($\Delta>0$), where the system 
flows to the strong coupling limit at lower temperatures. 
%
%\item[d)] 
\\(\textit{d})
Obtained data can be used to model the behavior observed in experiments
on materials with spin chains doped with magnetic and nonmagnetic 
impurities.\cite{hess} 
%\end{itemize}
%
%
%
\section{acknowledgments}
This work was supported by the FP6-032980-2 NOVMAG project and 
by the Slovenian Agency grant No. P1-0044.
\appendix
\section{Open chain}
\begin{figure}
\includegraphics[angle=0, width=.9\linewidth]{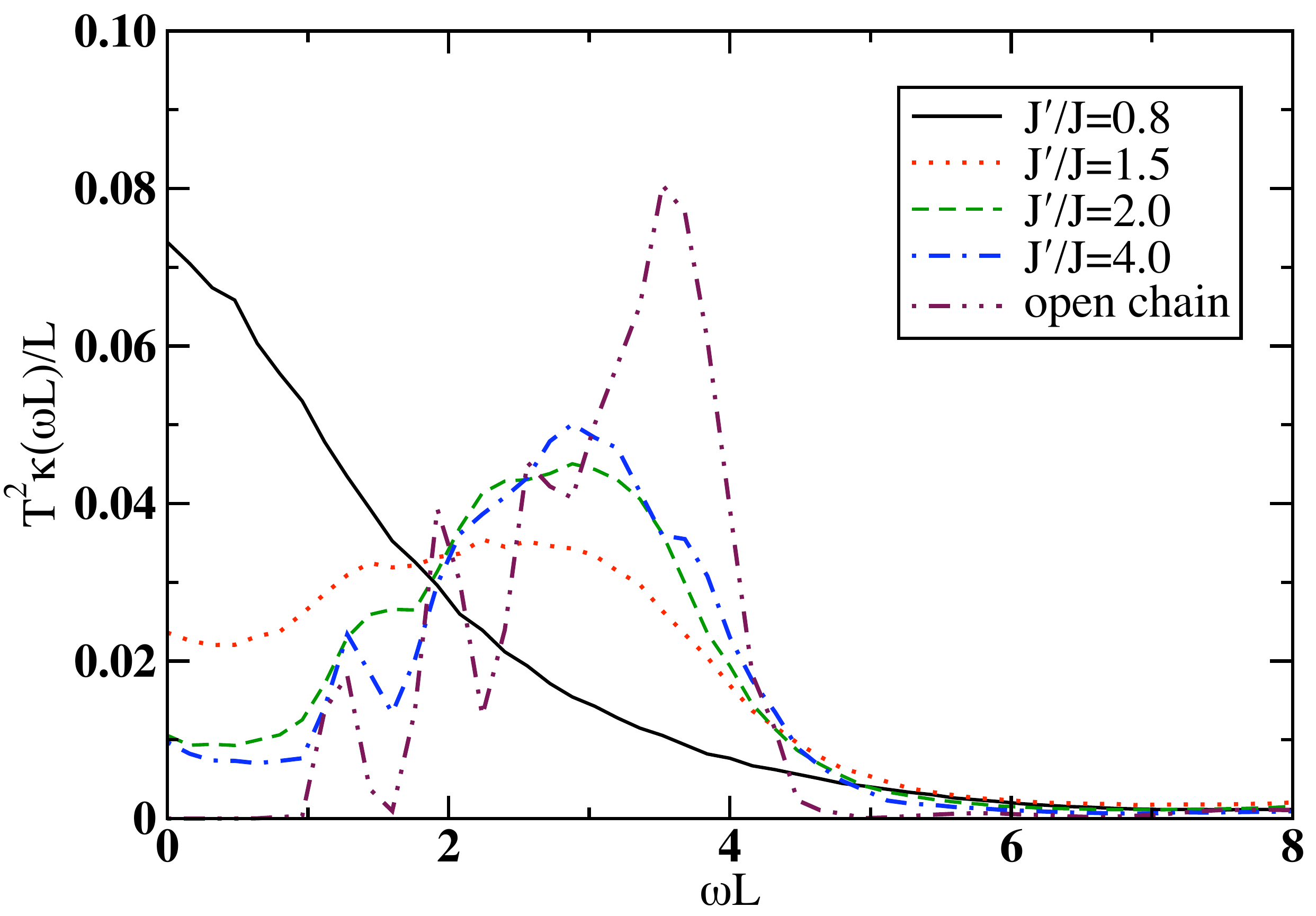}
\caption{
Frequency dependence of the thermal conductivity $\kappa(\omega L)/L$ 
in the high  temperature limit for various values of the coupling 
$J'/J=0.8-4.0$ and $\Delta=1.0$.}
\label{cutchain}
\end{figure}

Throughout the article the term ``cutting'' is used 
to describe the behavior of the system in the strong coupling limit. 
In order to justify the term ``cutting'', we present in Fig. \ref{cutchain} 
results for the thermal conductivity of a chain of  $L=16$  sites obtained 
by ED in the high temperature limit  for various couplings $J'$
and the thermal conductivity of a uniform chain with open boundary 
conditions as well. 
Fig. \ref{cutchain}  
illustrates the flow of the system from a Drude like behavior (weak coupling) 
to a chain with open boundary conditions (strong coupling), which
was already proposed for a single non-magnetic 
impurity (a local field) from the level statistics analysis.\cite{static} 
We choose to present the jagged results, i.e. without implementing any 
smoothing procedure, in order not to wash out the development of the 
narrow peaks corresponding  to the excitations of the open chain. 
For the strong coupling cases there is some rather significant structure
at frequencies $\sim J'$ which correspond to local excitations of the impurity.
However, these excitations are irrelevant for the effect of the impurity 
on the chain which is studied here.


\begin{references}
\bibitem{higherspin} It should be noted that a tower of 
integrable Hamiltonians exist for every value of spin, 
where the energy current is a conserved quantity, but 
these Hamiltonians have no obvious physical realizations.
\end{references}
\end{document}